# La-doping effect on spin-orbit coupled $Sr_2IrO_4$ probed by x-ray absorption spectroscopy

Jie Cheng[1*], Xuanyong Sun[2], Shengli Liu[1,3*], Bin Li[1*], Haiyun Wang[1], Peng Dong[4], Yu Wang[5] and Wei Xu[6,7]

[1] College of Science, Center of Advanced Functional Ceramics, Nanjing University of Posts and Telecommunications, Nanjing, Jiangsu 210023, China

[2] College of Electronic Science and Engineering, Nanjing University of Posts and Telecommunications, Nanjing, Jiangsu 210023, China

[3] Nanjing University (Suzhou) High-Tech Institute, Suzhou, 215123, China

[4] Information Construction and Management Office, Nanjing University of Posts and Telecommunications, Nanjing, Jiangsu 210023, China

[5] Shanghai Synchrotron Radiation Facility, Shanghai Institute of Applied Physics, Chinese Academy of Sciences, Shanghai 201204, China

[6] Beijing Synchrotron Radiation Facility, Institute of High Energy Physics, Chinese Academy of Sciences, Beijing 100049, China

[7] Rome International Center for Materials Science, Superstripes, RICMASS, via dei Sabelli 119A, I-00185 Roma, Italy

E-mail: chengj@njupt.edu.cn, liusl@njupt.edu.cn and libin@njupt.edu.cn

**Abstract.** $Sr_2IrO_4$ was predicted to be an unconventional superconductor upon carrier doping since it highly resembles the high-temperature cuprates. Here, to understand carrier doping effect on spin-orbit coupled Mott insulator $Sr_2IrO_4$, the electronic structure and local structure distortion for $Sr_{2-x}La_xIrO_4$ system have been investigated by x-ray absorption spectroscopy (XAS). By comparing the intensity of white-line features at the Ir $L_{2,3}$ absorption edges, we observe remarkably large branching ratios in La-doped compounds, greater than that of the parent material $Sr_2IrO_4$, suggesting a strong spin-orbit interaction (SOI) for $Sr_2IrO_4$-based system. Moreover, extended x-ray absorption fine structure (EXAFS) spectra demonstrate more regular $IrO_6$ octahedra, *i.e.* the weakened crystal electric field (CEF) *versus* La-doping. By theoretical calculations, the synergistic effect of regular $IrO_6$ octahedra and electron doping is established, which accounts for the transition from a Mott insulator to a conductive state in $Sr_{2-x}La_xIrO_4$-based system.

## 1. Introduction

Over the past quarter century, transition metal oxides (TMOs) have attracted vast attention in condensed matter due to the rich variety of unconventional physical behaviors including high-Tc superconductivity in cuprates [1], colossal magnetoresistance in manganates [2], and multiferroicity in Bismuth compounds [3]. Currently, the focus of many researchers has turned to Ir-based 5$d$ TMOs because of the observation of localized transport and magnetism [4, 5]. The strong spin-orbit interaction (SOI), which is proportional to $Z^4$ ($Z$ is the atomic number), plays a significant role in the iridates [6] and rigorously competes with other relevant energies, particularly the on-site Coulomb repulsion and crystal electric field (CEF) interaction arising from surrounding oxygen atoms in a nearly octahedral local coordinate environment [7]. Therefore, a new balance between the competing energies is established and drives exotic quantum phases in the iridates, such as the SOI - induced $J_{eff}$ = 1/2 Mott insulator [5].

Among all the iridates, the layered perovskite $Sr_2IrO_4$ known as SOI-related Mott insulator has been subjected to the most extensive investigations since it highly resembles the high-Tc cuprates in crystal structure, electronic structure and magnetic coupling constants [8]. $Sr_2IrO_4$, which has the $K_2NiF_4$ structure, is based on $Ir^{4+}$ in $IrO_6$ octahedra, where the electron configuration is [Xe]5$d^5$. The octahedral CEF splits the 5$d$ band states into $t_{2g}$ and $e_g$ states, and strong SOI split the $e_g$ manifold into two effective angular momentum energy levels: a doublet $J_{eff}$ = 1/2 and a quartet $J_{eff}$ = 3/2 [5]. For $Ir^{4+}$, the $J_{eff}$ = 3/2 bands are completely filled while the $J_{eff}$ = 1/2 bands are half filled. The presence of Coulomb repulsion, though small, further splits the narrow $J_{eff}$ = 1/2 bands into upper and lower Hubbard bands, which renders an insulating state [4].

Most interestingly, several theoretical investigations predicted that high-temperature superconductivity could be induced by carrier doping in this $Sr_2IrO_4$-based system [9-12]. Searching for novel superconductors in these SOI-induced Mott insulators is so meaningful that may stimulate a new surge of interest in superconductivity. Therefore, many experiments have been carried out to discover the possible novel superconductivity in electron and hole doped $Sr_2IrO_4$. For instance, a robust metallic state can be found in $La^{3+}$ doped $Sr_2IrO_4$ single crystal [13], while no metallic behavior is reported in the doped polycrystalline compounds [14]. Hole doping is performed by Sr vacancy in $Sr_2IrO_4$, where a semiconducting state is realized [15]. In addition, a combined experimental and theoretical study demonstrated that Rh substitution for Ir can lead to metallic behavior [16]. Very recently, experiments on doped $Sr_2IrO_4$ have uncovered tantalizing evidence of a Fermi surface split up into disconnected segments ("Fermi arcs") [17] and a low-temperature gap with d-wave symmetry [18, 19], which are hallmarks of the doped cuprates [20]. To the best of our knowledge, the novel superconductivity however has not been found yet in doped $Sr_2IrO_4$ system. As a consequence, a more accurate and systematic investigation of carrier doping effect is still necessary to better understand the $Sr_2IrO_4$ system, that might be beneficial to uncover the potential superconductivity in SOI-related Mott insulators.

As is known to all, material properties are in a close relationship with its atomic structure. For example, data pointed out that carrier doping can alter lattice parameters and further change the electronic and magnetic behaviors of $Sr_2IrO_4$ [14, 15, 21]. Furthermore, Bhatti *et al.* established a relationship between temperature dependent magnetic property and the structural parameters by means of x-ray diffraction (XRD) measurement on $Sr_2IrO_4$ [22]. Synchrotron-radiation-based x-ray absorption spectroscopy (XAS), however, is a well-recognized local experimental technique capable of investigating the local atomic and electronic structure of a material [23]. Moreover, Ir $L_{2,3}$-edge XAS can be competent for providing valuable information about the strength of the SOI in Ir-based systems [24]. In this contribution, we take the polycrystalline compounds of La-doped $Sr_2IrO_4$ as the object of our research, and report detailed investigations of La-doping effect, through a comprehensive analysis of Ir $L_{2,3}$-edge XAS.

## 2. Experiments and Calculations

Polycrystalline samples of $Sr_{2-x}La_xIrO_4$ ($x$ = 0, 0.1, 0.2, 0.3) were synthesized through the conventional solid-state reaction method as mentioned elsewhere [25]. We collected both the Ir $L_2$-edge x-ray absorption near edge structure (XANES) and $L_3$-edge extended x-ray absorption fine structure (EXAFS) spectra of $Sr_{2-x}La_xIrO_4$ at room temperature, which occur at energy of 12.824keV and 11.215keV, respectively. XAS measurements were performed in transmission mode using the BL-14W1 beamline of Shanghai Synchrotron Radiation Facility (SSRF). Several scans were collected to ensure the spectral reproducibility. The storage ring was working at electron energy of 3.5 GeV, and the maximum stored current was about 250mA. Data were recorded using a Si (111) double crystal monochromator and normalized by the IFEFFIT program package [26].

Density functional theory (DFT) calculations for electron-doped $Sr_2IrO_4$ system were carried out within the local-density approximation (LDA) including SOI and on-site Coulomb interactions using OPENMX [27], based on the linear-combination-of-pseudo-atomic-orbitals (LCPAO) method [28]. SOI is treated via a fully relativistic j-dependent pseudopotential in the non-collinear DFT scheme, with 300 Ry of energy cutoff and $11\times11\times11$ k-grid. For doped $Sr_2IrO_4$, only one of the eight Sr atoms was replaced by La in a unit cell, which gives atomic ratio of La *versus* Sr as 0.125:0.875. The on-site Coulomb interaction parameter $U_{eff}$, is set to 2.0 eV for Ir d-orbitals in our LDA + SOI + U calculations.

## 3. Discussions and Results

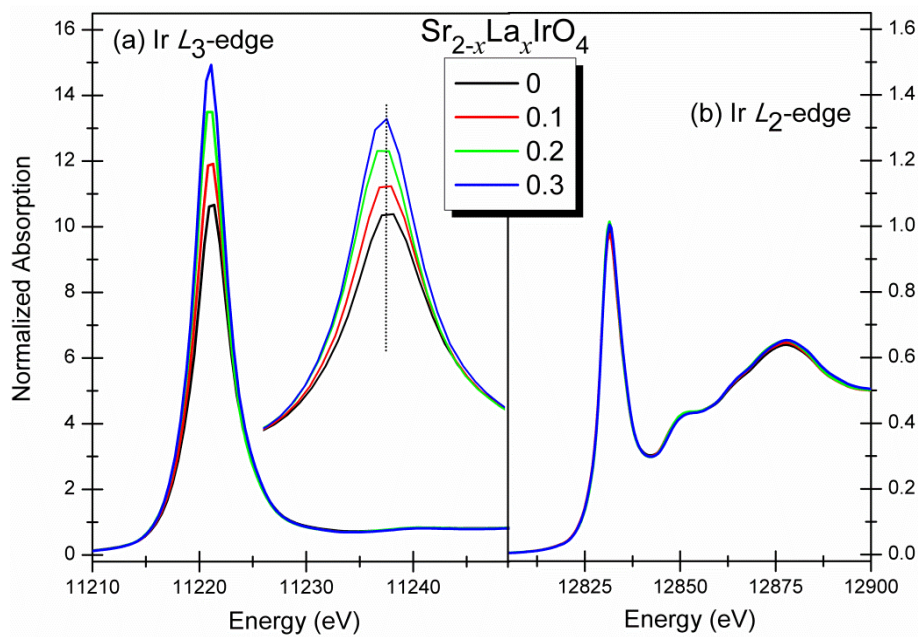

**Fig.1** Ir $L_{2,3}$-edge normalized XANES spectra of $Sr_{2-x}La_xIrO_4$. The $L_2$-edge spectra are normalized at half of the $L_3$-edge step size. The "white-line" (WL) feature at Ir $L_3$-edge is magnified in the inset.

Previous electrochemical reports conjecture the presence of $Ir^{3+}$ in $Sr_{2-x}La_xIrO_4$ system [14]. Here, to obtain information on the chemical valance of Ir in La-doped $Sr_2IrO_4$ system, we display the Ir $L_{2,3}$-edge XANES spectra of $Sr_{2-x}La_xIrO_4$ compounds in Fig. 1. The most striking characteristic of the Ir $L_{2,3}$-edge XANES spectra are the sharp "white-line" (WL) feature that corresponds to $2p \rightarrow 5d$ electronic transitions. From the dipole selection rules we know that $\Delta J$ must be equal to 0 or ±1 so the Ir $L_2$-edge is sensitive to transitions involving $5d_{3/2}$ (i.e. $J_{eff} = 3/2$) holes, while the $L_3$-edge is related to both $5d_{5/2}$ (upper Hubbard band associated with $J_{eff} = 1/2$ and the CEF $e_g$ manifolds) and $5d_{3/2}$ final states. Because the WL features are more pronounced at the $L_3$ edge than the $L_2$ edge, we can infer that these unoccupied 5d states are primarily $5d_{5/2}$ rather than $5d_{3/2}$ in nature, consistent with the almost completely filled $J_{eff} = 3/2$ bands. From Fig. 1, both a broadening of the Ir $L_3$-edge WL peak and a negative shift in spectral weight with increasing La concentration may be some indications of the presence of $Ir^{3+}$ introduced by La doping. However, conventional Ir $L_{2,3}$-edge XANES cannot effectively observe changes of the chemical valence of Ir due to the effect of core-hole lifetime broadening. By suppressing such core-hole lifetime effects, partial fluorescence yield (PFY)-XAS can provide a significant improvement in experimental energy resolution [29], and in our future work, PFY-XAS technique will be further utilized to confirm our conclusion.

In essence, of great significance in the study of $Sr_2IrO_4$-based materials is the SOI. Recent work has pointed out that the SOI could be reflected in the measurement of the branching ratio, $BR = I_{L_3}/I_{L_2}$, where $I_{L_{2,3}}$ is the integrated intensity of the WL feature at the $L_2$ and $L_3$ edges, respectively [24]. Using theoretical arguments proposed by van der Laan and Thole [30, 31], the branching ratio can be related to the SOI via $BR = (2+r)/(1-r)$ with $r = \langle L \cdot S \rangle / n_h$, where $n_h$ refers to the number of holes on the transition-metal site and $\langle L \cdot S \rangle$ is the expectation value of the SOI of the empty states. It has already been shown that the BR ~ 2 corresponds to a negligible SOI, and BR > 2 if $\langle L \cdot S \rangle \neq 0$ [30, 31]. Thus we investigate the SOI through the branching ratio of $Sr_2IrO_4$-based compounds.

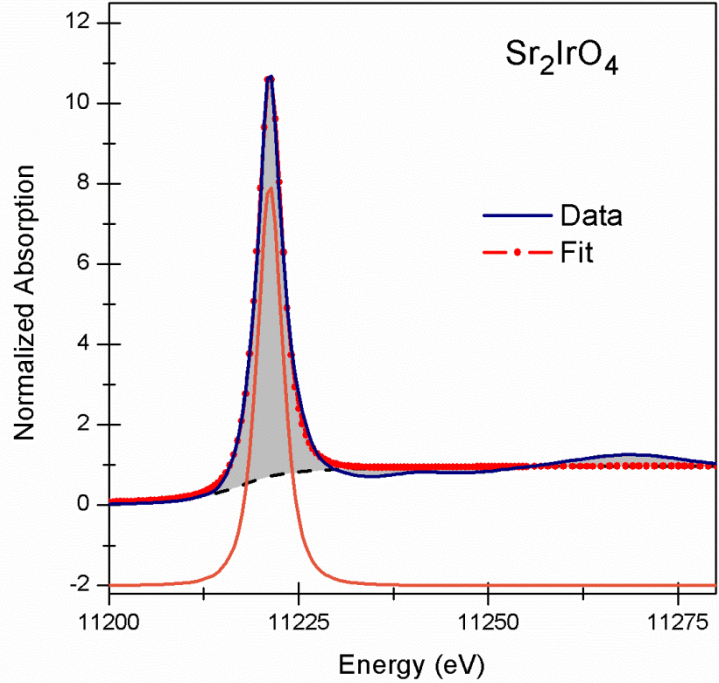

**Fig. 2** Experimental determination of the white-line intensity at the Ir $L_3$-edge XANES in $Sr_2IrO_4$. The solid blue line represents the experimental XANES, while the solid red line + symbol is the best fit to the data using a Lorentzian + arctangent fit function. The dashed line is an arctangent background function. In the bottom the background-subtracted white line peak is shown.

As mentioned above, in order to discuss the SOI effect on a quantitative level, it is necessary to accurately determine the integrated intensity of the WL features at the $L_2$ and $L_3$ edges. Here, the Ir $L_{2,3}$-edge XANES spectra are fitted based on the least-squares method using a Lorentzian + arctangent fit function, illustrated in Fig. 2. The continuum edge-step is modelled by an arctangent function, and we fit the WL feature using a Lorentzian function. The only fixed parameter in these fits was the amplitude of the arctangent function, which was set to unity at the $L_3$ edge (0.5 at the $L_2$ edge) in order to match the continuum edge step. We found BR = 6.46, which is greater than the statistical one, suggesting a strong SOI effect in parent material $Sr_2IrO_4$. With $n_h = 5$, we obtain $\langle L \cdot S \rangle = 2.99$ in units of $\hbar^2$, consistent with other reports [30].

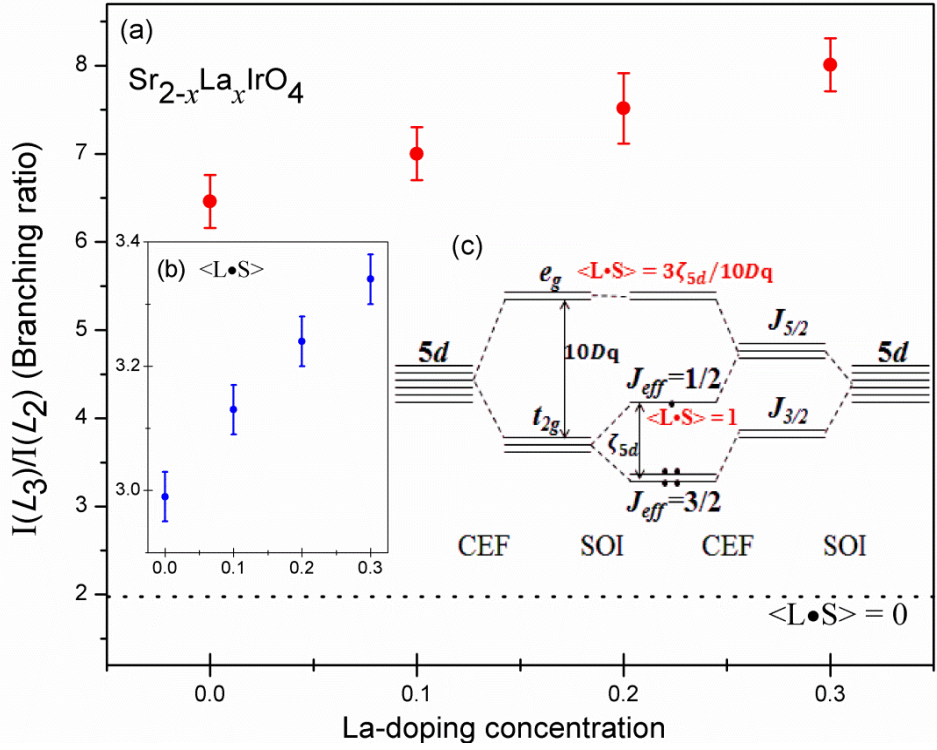

**Fig. 3** The branching ratio $I_{L_3}/I_{L_2}$ (a) and $\langle L \cdot S \rangle$ (b) as a function of La-doping concentration. The dashed line represents the statistical branching ratio of $I_{L_3}/I_{L_2} = 2$ obtained in the limit of negligible SOI. (c) A schematic energy diagram of $5d$ level splittings by the CEF and SOI, based on Ref. 4.

Upon La-doping we observe significantly increased branching ratio and $\langle L \cdot S \rangle$, as shown in Fig. 3. With increasing La-doping concentration, the branching ratio of $Sr_{2-x}La_xIrO_4$ increases from 6.46 ($x = 0$) to 8.01 ($x = 0.3$), and the derived $\langle L \cdot S \rangle$ increases from 2.99 to 3.34 $\hbar^2$, respectively. Since $\langle L \cdot S \rangle$ is a property of the local moment, it is mostly determined by the SOI and the CEF acting on $5d$ electrons [32]. Note that since XANES probes all empty $5d$ states, the measured $\langle L \cdot S \rangle$ will be the sum of two contributions (illustrated in Fig. 3(c)): a single hole in the $J_{eff} = 1/2$ state ($\langle L \cdot S \rangle = 1$) [33] and four holes in the $e_g$-derived states ($\langle L \cdot S \rangle = 4 \times 3\zeta_{5d}/10Dq$) [34], where $\zeta_{5d}$ represents the SOI of the $5d$ states and $10Dq$ is the parameter of octahedral CEF. Therefore, for $Sr_2IrO_4$-based system La-doping can effectively alter the relative strength of the SOI and CEF.

As we all know, the Ir atoms make a significant contribute to the strong SOI in $Sr_2IrO_4$-based system, and the SOI strength changes due to the altering environment, *e.g.* the cation substitutions [35]. For example, an isoelectronic substitution of Ir with Rh in $Sr_2Ir_{1-x}Rh_xO_4$ would tune the SOI strength from the strong 5$d$ regime to the moderate 4$d$ regime [35]. However, XAS technique clearly demonstrated that Rh-doping introduces $Rh^{3+}$ and $Ir^{5+}$ ions into the material, which not only tune SOI but also alter the band filling *via* hole doping [36]. Similarly, a substitution of 5$d$-$Ir^{4+}$ with 4$d$-$Ru^{4+}$ in $Sr_2Ir_{1-x}Ru_xO_4$ simultaneously tunes the band filling (hole doping) as well as SOI [35, 37]. Combined with the above two results, the substitution at Ir sites can significantly tune the SOI strength of the system. On the other hand, the substitution of divalent Sr by trivalent La in $Sr_{2-x}La_xIrO_4$ just tunes the band filling (electron doping) [35]. Roughly speaking, La-doping imposes negligible effect on the SOI strength, and we neglect the SOI changes in our discussion. Then we will focus on the CEF effect *versus* La-doping concentration.

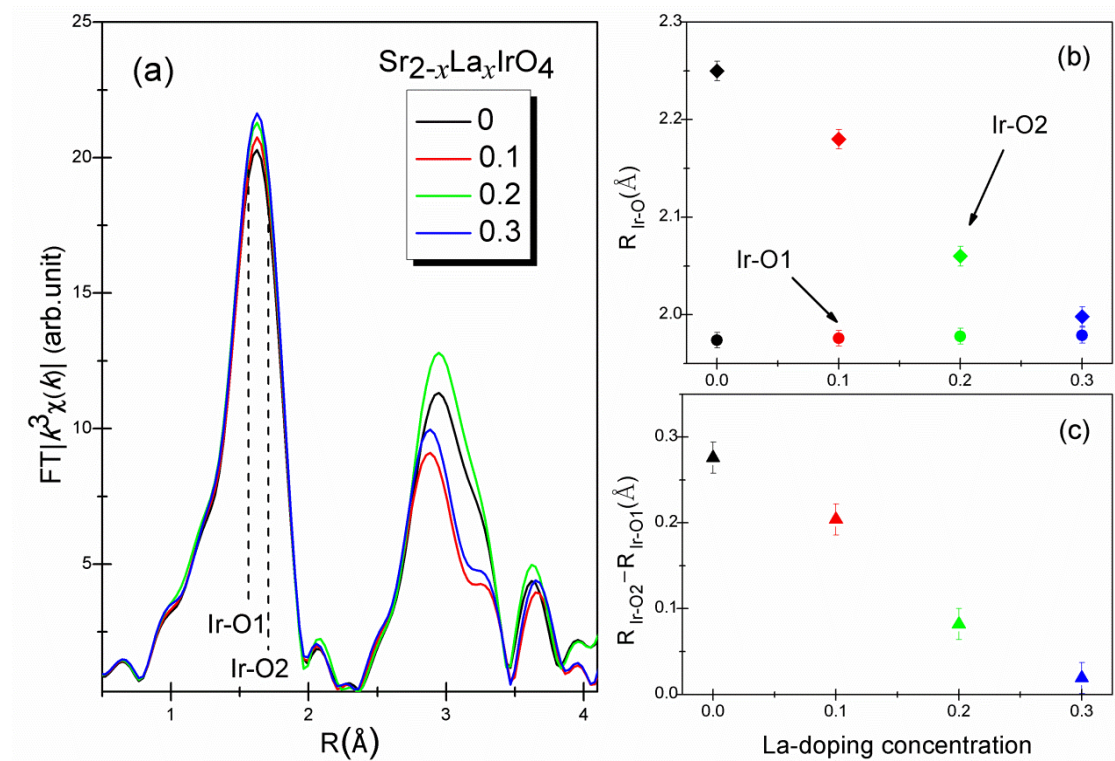

**Fig. 4** (a) Fourier transforms of the $k^3\chi(k)$ EXAFS signals of $Sr_{2-x}La_xIrO_4$ compounds at Ir $L_3$-edge. Results from the fit of Ir-O1/Ir-O2 bond distances are shown in panel (b). (c) The derived differences between the distances of Ir-O1 and Ir-O2 bonds as a function of La-doping concentration. Error bars correlated to the uncertainty value associated to the EXAFS analysis are given.

For $Sr_2IrO_4$-based system, the CEF interaction originates from surrounding six oxygen atoms around Ir in a nearly octahedral environment [7]. The $IrO_6$ octahedron has two distinct oxygen positions (four basal O1 and two apical O2) where they are found to be slightly elongated apically. Therefore, to gain an insight into the structural distortion induced by La-doping, we have undertaken detailed structural study (e.g. the bond distances of Ir-O1 and Ir-O2) by means of Ir $L_3$-edge EXAFS measurements. The Fourier transform (FT) of the EXAFS signal as a function of La-doping is displayed in Fig. 4(a). We have to underline that the positions of the peaks in the FT are shifted a few tenths of Å from the actual interatomic distances because of the EXAFS phase shift [38]. The first peak in the FTs includes two sub-shells around the Ir atom, i.e. four nearest basal O1 and two apical O2. In order to obtain quantitative results, we fit the first peak of EXAFS spectra which were isolated from the FTs with a rectangular window. The range in $k$ space was 2 ~ 15 Å$^{-1}$ and that in $R$ space was 1.0 ~ 2.2 Å. The number of independent parameters which could be determined by EXAFS is limited by the number of the independent data points $N_{ind} \sim (2\Delta k \Delta R)/\pi$, where $\Delta k$ and $\Delta R$ are respectively the ranges of the fit in the $k$ and $R$ space. Thus $N_{ind}$ is 10 ($\Delta k$ = 13 Å$^{-1}$, $\Delta R$ = 1.2 Å), sufficient to obtain all parameters. Results in Fig. 4(b) point out a slow increase in basal Ir-O1 distance, while the apical Ir-O2 distance decreases dramatically *versus* La-doping, generally in line with the results of pervious Rietveld refinements of XRD [14, 25]. As presented in Fig. 4(c) La dopants make a significant contribution to reduce the distance difference between Ir-O1 and Ir-O2 bonds. Thus we found more regular $IrO_6$ octahedra upon La-doping, which could subsequently weaken the CEF.

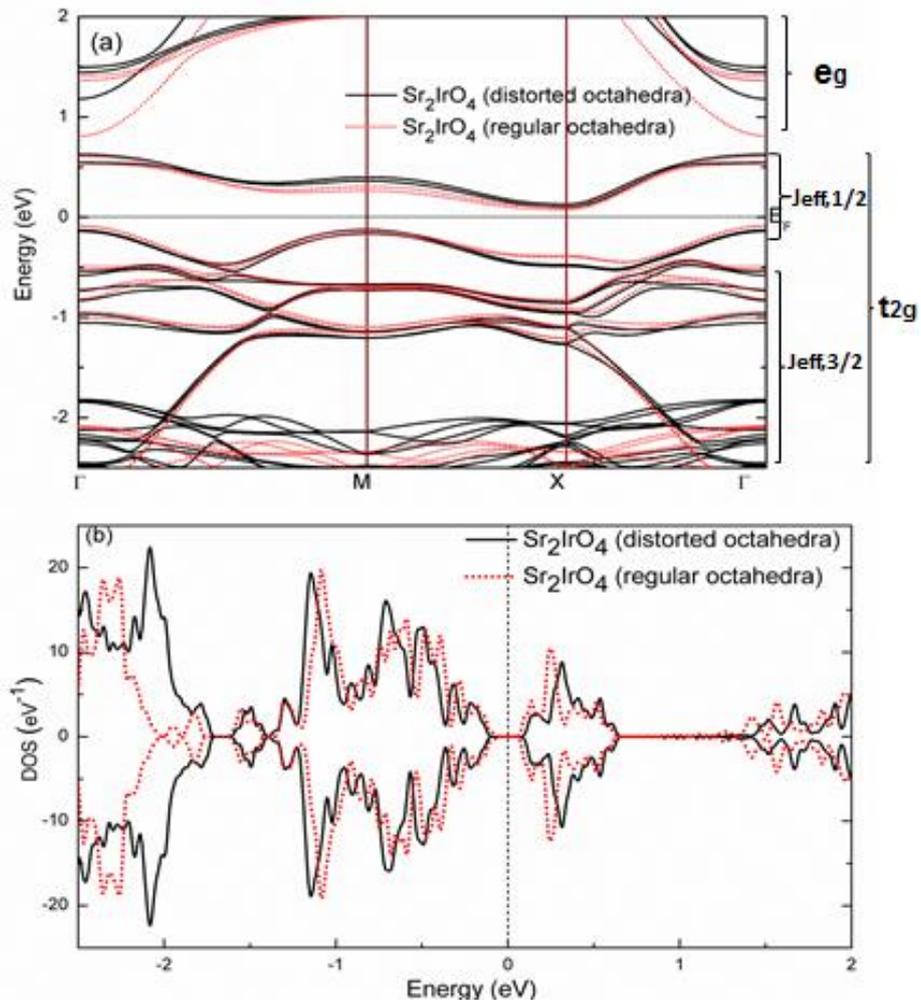

**Fig. 5** The comparison of band structures (a) and DOS (b) for $Sr_2IrO_4$ with distorted octahedra and regular octahedra. All energies are relative to the Fermi energy.

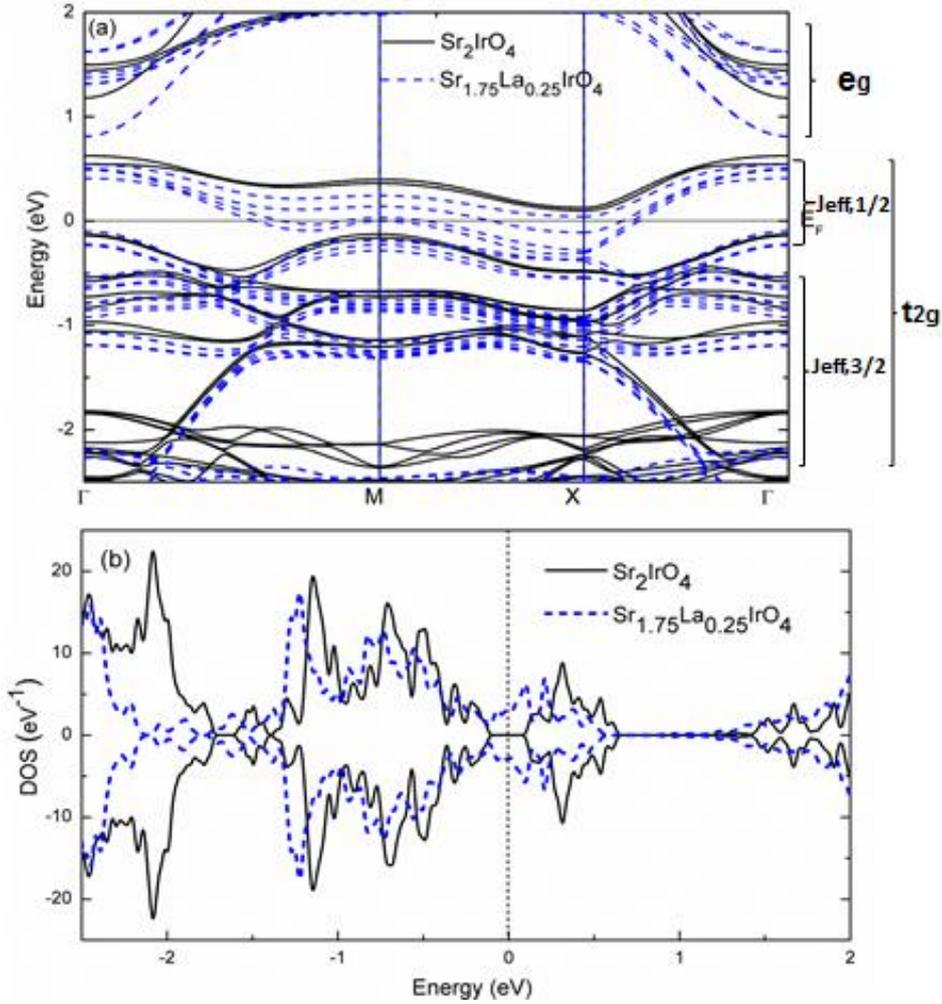

**Fig. 6** Comparison of the band structures (a) and DOS (b) for $Sr_2IrO_4$ and $Sr_{1.75}La_{0.25}IrO_4$ (with regular octahedra).

In order to investigate both the regular octahedra and La-doping effect on band structures, we performed DFT calculations on $Sr_2IrO_4$ with distorted octahedra, regular octahedra and $Sr_{1.75}La_{0.25}IrO_4$ with regular octahedra. From Fig. 5, we found that the band gap between upper and lower Hubbard bands (*i.e.* the splits of $J_{eff} = 1/2$ states) of regular octahedra becomes slightly decreased, with respect to that of distorted octahedra. Note that all the potential parameters are fixed to perform all these calculations. Hence it is assured that the decreased band gap is attributed to the regular octahedra. Furthermore, the regular octahedra can effectively shift down the $e_g$ states, which lead to a reduction of $10Dq$ from 2.0eV of distorted octahedra to 1.7eV of regular octahedra. This decrease of $10Dq$ is sufficient to cause the enhancement of <L S> by more than 10%, in agreement with the experimental results of

XANES. These results demonstrate that the regular octahedra are crucial to $Sr_2IrO_4$-based system, which effectively tune the band gap and the CEF.

Based on the regular $IrO_6$ octahedra, the La impurity (electron doping) tunes the band filling, shifts up the $E_F$ and turns off the band gap, promoting the emergence of a conductive state (in Fig. 6). Therefore, it is the synergistic effect of regular $IrO_6$ octahedra and electron doping that results in the transition from a Mott insulator to a conductive state in $Sr_{2-x}La_xIrO_4$-based system.

## 4. Conclusion

To summarize, we have investigated the La-doping effect on the spin-orbit coupled $Sr_2IrO_4$ probed by Ir $L_{2,3}$-edge XAS. First of all, we found the presence of $Ir^{3+}$ introduced by La doping, but further PFY-XAS technique will be utilized to confirm it. By fitting the white line features of Ir $L_{2,3}$-edge XANES, a remarkably large branching ratio is observed, suggesting a strong SOI effect in parent material $Sr_2IrO_4$. Upon La-doping, this branching ratio and the derived ⟨L•S⟩ increase gradually. For $Sr_2IrO_4$-based system, the value of ⟨L•S⟩ is intimately related to the ratio of SOI and CEF. Ir $L_3$-edge EXAFS data pointed out a dramatic decrease of the apical Ir-O2 distance, *i.e.* more regular $IrO_6$ octahedra upon La-doping, which inevitably weakens the CEF. Finally, theoretical calculations suggest the synergistic effect of regular $IrO_6$ octahedra and electron doping that leads to a conductive state in $Sr_{2-x}La_xIrO_4$-based system.

**Acknowledgement**

This work was partly supported by the National Natural Science Foundation of China (NSFC 11405089 and 11504182), the Natural Science Foundation of Jiangsu Province of China (No. BK20130855 and 20150831), the "Six Talents Peak" Foundation of Jiangsu Province (2014-XCL-015), the Nanotechnology Foundation of Suzhou Bureau of Science and Technology (ZXG201444) and the Scientific Research Foundation of Nanjing University of Posts and Telecommunications (No. NY213053).